\pdfoutput=1

\documentclass[reprint,amsmath,amssymb,floatfix,aps,longbibliography]{revtex4-1}
\usepackage{graphics,graphicx,float}

\usepackage[pdftex,
            pdfauthor={Steven A. Frank},
            pdftitle={},
            pdfsubject={Natural selection. V. How to read the fundamental equations of evolutionary change in terms of information theory}]{hyperref} 

\def\citeyear{\citep}
\def\autocite{\citep}

\newcommand{\zbar}{\bar{z}}
\newcommand{\wbar}{\bar{w}}
\newcommand{\mbar}{\bar{m}}
\newcommand{\cov}{{\hbox{\rm Cov}}}

\newcommand{\KL}[2]{\mathcal D\left(#1||#2\right)}
\newcommand{\D}{\mathcal D}
\newcommand{\J}{J}
\newcommand{\qfrac}{\frac{q_i'}{q_i}}
\newcommand{\qfraclog}{\log\left(\frac{q_i'}{q_i}\right)}
\newcommand{\dlog}{\GD\log(q_i)}
\newcommand{\GDs}{\GD_s}
\newcommand{\GDc}{\GD_c}
\newcommand{\qdot}{\dot{q}}

\newcommand{\Gb}{\beta}

\newcommand{\GD}{\Delta}

\newcommand{\Gm}{\mu}

\newcommand{\Gth}{\theta}

\newcommand{\dd}{{\hbox{\rm d}}}

\newcommand{\Eq}[1]{Eq.~(\ref{eq:#1})}

\newcommand{\boldrule}{\hrule height 1.2pt}
\newcommand{\noterule}{\medskip\boldrule\medskip}	

\newcommand{\noterulenote}[1]{\null}

\newcount\BoxNum \BoxNum 1\relax
\makeatletter
\newcommand{\boxlabel}[1]{%
  \protected@write \@auxout {}{\string \newlabel {box:#1}{{\the\BoxNum}{\thepage}{\noexpand\relax}%
  	{\@ifundefined{hyper@@anchor}{\relax}{box.\the\BoxNum}}%
  	{}}}%
  \@ifundefined{hyper@@anchor}{}{\hypertarget{box.\the\BoxNum}{}}%
  \advance\BoxNum 1\relax}
\makeatother
\newcommand{\Boxx}[1]{Box~\ref{box:#1}}
\newcommand{\BoxLabel}{Box~\the\BoxNum}

\begin{document}

\title{Natural selection. V. How to read the fundamental equations of evolutionary change in terms of information theory}

\author{Steven A.\ Frank}
\email[email: ]{safrank@uci.edu}
\homepage[homepage: ]{http://stevefrank.org}
\affiliation{Department of Ecology and Evolutionary Biology, University of California, Irvine, CA 92697--2525  USA}

\begin{abstract}

\noindent The equations of evolutionary change by natural selection are commonly expressed in statistical terms.  Fisher's fundamental theorem emphasizes the variance in fitness. Quantitative genetics expresses selection with covariances and regressions.  Population genetic equations depend on genetic variances.  How can we read those statistical expressions with respect to the meaning of natural selection?  One possibility is to relate the statistical expressions to the amount of information that populations accumulate by selection.  However, the connection between selection and information theory has never been compelling.  Here, I show the correct relations between statistical expressions for selection and information theory expressions for selection.  Those relations link selection to the fundamental concepts of entropy and information in the theories of physics, statistics, and communication.  We can now read the equations of selection in terms of their natural meaning.  Selection causes populations to accumulate information about the environment\footnote{\href{http://dx.doi.org/10.1111/jeb.12010}{doi:\ 10.1111/jeb.12010} in \textit{J. Evol. Biol.}}\footnote{Part of the Topics in Natural Selection series. See \Boxx{preface}.}.

\end{abstract}

\maketitle

\begin{quote}
There are difficulties in applying information theory in genetics. They arise principally, not in the transmission of information, but in its meaning \autocite[p.~181]{maynard-smith00the-concept}.
\end{quote}

\section*{Introduction}

I show that natural selection can be described by the same measure of information that provides the conceptual foundations of physics, statistics and communication.  Briefly, the argument runs as follows.  The classical models of selection express evolutionary rates in proportion to the variance in fitness. The variance in fitness is equivalent to a symmetric form of the Kullback-Leibler information that the population acquires about the environment through the changes in gene frequency caused by selection. 

Kullback-Leibler information is closely related to Fisher information, likelihood, and Bayesian updating from statistics, as well as Shannon information and the measures of entropy that arise as the fundamental quantities of communication theory and physics.  Thus, the common variances and covariances of evolutionary models are equivalent to the fundamental measures of information that arise in many different fields of study.

In Fisher's fundamental theorem of natural selection, the rate of increase in fitness caused by natural selection is equal to the genetic variance in fitness.  Equivalently, the rate of increase in fitness is proportional to the amount of information that the population acquires about the environment \autocite{frank09natural}. 

In my view, information is a primary quantity with intuitive meaning in the study of selection, whereas the genetic variance just happens to be an algebraic equivalence for the measure of information.  The history of evolutionary theory has it backwards, using statistical expressions of variances and covariances in place of the equivalent and more meaningful expressions of information.  To read the fundamental equations of evolutionary change, one must learn to interpret the standard expressions of variances and covariances as expressions of information.  

\section*{Overview}

The first section reviews the classic statistical expressions for selection.  Evolutionary change caused by selection is the covariance between fitness and character value.  That covariance equals the regression of character value on fitness multiplied by the variance in fitness.

The second section expresses selection in terms of the classic equations from information theory (see \Boxx{info}).  I show that the change in the mean logarithm of fitness is the Jeffreys information divergence.  That divergence measures the accumulation of information by natural selection between the initial population and the population after it has been updated by selection.  The relations between the statistical and information perspectives follow by connecting the classic statistical expressions of selection to the new information description for selection.

The third section analyzes the Jeffreys divergence as the measure of information in the fundamental equations of selection.  The Jeffreys divergence is the sum of two expressions for relative entropy. Relative entropy, known as the Kullback-Leibler divergence, measures the gain in information with regard to an abstract and universal notion of encoding, independently of the meaning of that information.  A universal, abstract measure of information in terms of encoding allows a general theory of information to provide the foundation for the deepest concepts in communication, physics and statistics.

The fourth section concerns the meaning of information.  Although encoding provides a useful measure with 

\begin{figure}[H]
\begin{minipage}{\hsize}
\parindent=15pt
\noterule
{\bf \noindent\BoxLabel. Topics in the theory of natural selection}
\noterule
\noindent This article is part of a series on natural selection.  Although the theory of natural selection is simple, it remains endlessly contentious and difficult to apply.  My goal is to make more accessible the concepts that are so important, yet either mostly unknown or widely misunderstood.  I write in a nontechnical style, showing the key equations and results rather than providing full derivations or discussions of mathematical problems.  Boxes list technical issues and brief summaries of the literature.    
\noterule
\end{minipage}
\end{figure}
\boxlabel{preface}

\noindent regard to information theory, we must also interpret the meaning of that information in terms of selection.  Meaning arises by the relation of encoded information to whatever scale we use to interpret a particular problem.  For selection, we interpret meaning with regard to characters.  Characters may be gene frequencies or measurements made on individuals. Characters lead to a general notion of the scale for meaning with respect to the scale of encoded information.

The fifth section explicitly connects the abstract scale of encoded information to the meaningful scale of information in problems of selection.  The analysis leads to the relation between the Jeffreys divergence, the most general expression for selection, and Fisher information as the limiting form of the Jeffreys divergence when changes in magnitude are small.  Fisher information is the sensitivity of changes in abstract encoded information relative to the distance that one moves along a scale of meaning.  Encoded information is equivalent to the log-likelihood ratio, which is why Fisher information provides the conceptual foundations for the theory of statistics.

The sixth section uses Fisher information to derive various elegant expressions for selection.  For example, suppose that changes in the average value of a character sufficiently describe the changes caused by selection. Then mean log fitness increases by the Fisher information in an observation about the average character value multiplied by the squared change in the average character value.  This expression connects the scale of encoded information, which is mean log fitness, to the scale of meaning, which in this case is the average value of a character in the population.

The seventh section relates the parametric description of characters to a more general nonparametric expression.  In the previous example, the change caused by selection was described fully by a change in a parameter, the mean  In the general case, no parametric summary statistics fully capture the change in populations.  Instead, one must use the full range of different types in the population, providing a nonparametric description of the change in the distribution of frequencies caused by selection.  The full nonparametric expression shows the universal applicability of the equations selection and information.

\begin{figure}[H]
\begin{minipage}{\hsize}
\parindent=15pt
\noterule
{\bf \noindent\BoxLabel. Information, entropy and complexity}
\noterule
\noindent \textcite{cover91elements} give an excellent introduction to information theory and its applications.  \textcite{jaynes03probability} is a fascinating analysis of the connections between information, entropy, probability, Bayesian analysis, and statistical inference. \textcite{kullback59information} is a broad synthesis of information theory in relation to classical statistics.  Fisher's \citeyear{fisher22on-the-mathematical,fisher25theory} original papers on the theoretical foundations of statistics set the basis for all future work on information and statistics, with the 1925 paper showing the key role of Fisher information.  

Entropy arose in the study of thermodynamics \autocite{clausius67mechanical,boltzmann72weitere,gibbs02elementary}.   \textcite{ben-naim08entropy} gives a simple introduction. \textcite{hill87an-introduction} provides a classical text.  Information theory arose in Fisher's work and separately in the study of communication through the analyses of \textcite{hartley28transmission} and \textcite{shannon48a-mathematical-a,shannon48a-mathematical-b}. The underlying concepts of entropy and information are very close.  Some think the concepts are identical, but controversy remains \autocite{jaynes03probability,ben-naim08a-farewell}.

\textcite{jeffreys46an-invariant} divergence first appeared in an attempt to derive prior distributions for use in Bayesian analysis rather than as the sort of divergence used in this article.  \textcite{kullback51on-information} and \textcite{kullback59information} presented both the asymmetric divergence $\D$, given in \Eq{KLdef}, which is now known as the Kullback-Leibler divergence, and the symmetric form, $\J$, given in \Eq{J}, which is now known as the Jeffreys divergence.  They noted Jeffreys' previous usage of $\J$ in the context of Bayesian priors, and then developed the importance of the divergence interpretation for statistical theory, particularly the asymmetric form, $\D$.

I do not discuss Kolmogorov complexity in this article. However, it is an important concept that may ultimately prove as interesting for biological applications as the classic analyses of entropy and information.  Kolmogorov complexity measures the information content of an object (individual) by the shortest binary computer program that fully describes the object \autocite{cover91elements,li08an-introduction}.  At the population level, the average Kolmogorov complexity often has a close association with the formal theories of entropy and information, but it is not exactly the same.  

With respect to selection, fitness is, in essence, the match of characters to environmental challenge.  That match depends on the algorithmic relation between the information content of an organism and the interpretation of that information through the development of phenotype.  Development is not exactly like running a computer program encoded in the genes, but the analogy is not so far off.  I suspect that, someday, Kolmogorov complexity or related measures will help to understand biochemical, developmental and evolutionary processes.  A few authors have taken the first steps \autocite{gell-mann96information,adami00physical,adami02what}.    
\noterule
\end{minipage}
\end{figure}
\boxlabel{info}

The eighth section distinguishes changes by selection from total evolutionary change.  Numerous extrinsic and unpredictable forces beyond selection can change the characteristics of populations and their fit to the environment. I show the full expression for evolutionary change, placing selection in the broader evolutionary context.  No general conclusion about total evolutionary change is possible, because the complete range of forces that can perturb populations remains unpredictable.  However, we can express an elegant equilibrium condition.  At equilibrium, the gain in information by selection must be exactly balanced by the decay in information caused by other evolutionary forces.

The Discussion reviews the main argument.  Classic equations for selection describe change by statistical expressions of covariances, variances, and regressions.  In terms of encoded information, the change caused by selection is the Jeffreys divergence.  A generalized notion of Fisher information connects encoded information to the scale of meaning.  By equating the statistical description with the information description, we learn how to read the fundamental equations of selection in terms of information.

\section*{Classic equations of natural selection}

Equations of natural selection are often expressed in the statistical language of population variances, covariances, and regressions.  In this section, I show how these statistical expressions arise from the simplest models of selection. Later sections connect these classic equations to the amount of information that a population accumulates by selection.

Textbooks on population genetics and quantitative genetics present the classic equations of selection \autocite{crow70an-introduction,falconer96introduction,roff97evolutionary,futuyma98evolutionary,lynch98genetics,charlesworth10elements,ewens10mathematical}. Lande developed the statistical nature of selection equations \autocite{lande79quantitative,lande83the-measurement}, see also \textcite{frank97the-price}.

\subsection*{Selection}

A simple model starts with $n$ different types of individuals.  The frequency of each type is $q_i$.  Each type has $w_i$ offspring, where $w$ expresses fitness.  In the simplest case, each type is a clone producing $w_i$ copies of itself in each round of reproduction.  

The frequency of each type after selection is 
\begin{equation}\label{eq:replicator}
  q_i' = q_i\left(\frac{w_i}{\wbar}\right),
\end{equation}
where $\wbar = \sum q_iw_i$ is average fitness. The summation is over all of the $n$ different types indexed by the $i$ subscripts.  See \Boxx{price} for the proper interpretation of $q_i'$.

This equation is called a haploid model in classical population genetics, because it expresses the dynamics of different alleles at a haploid genetic locus.  Recently, economists, mathematicians, and game theorists have called this expression the replicator equation, because it expresses in the simplest way the dynamics of replication \autocite{taylor78evolutionary,hofbauer98evolutionary,hofbauer03evolutionary}.

It is often convenient to rewrite \Eq{replicator} as the change in the frequency of each type, $\GD q_i=q_i'-q_i$. Subtracting $q_i$ from both sides of \Eq{replicator} yields
\begin{equation}\label{eq:aveEx}
  \GD q_i = q_i\left(\frac{w_i}{\wbar} - 1\right).
\end{equation}
\Boxx{price} describes a universal interpretation of these equations for selection that transcends the narrow haploid and replicator models.

\subsection*{Characters}

Eqn \ref{eq:aveEx} describes change in frequency.  How does selection change the value of characters?  Suppose that each type, $i$, has an associated character value, $z_i$.  The average character value in the initial population is $\zbar = \sum q_iz_i$.  The average character value in the descendant population is $\zbar' = \sum q_i'z_i'$, where $z_i'$ is the character value in the descendants (see \Boxx{price}).  For now, assume that descendants have the same character value as their parents, $z_i' = z_i$.  Then $\zbar' = \sum q_i'z_i$, and the change in the average value of the character caused by selection is
\begin{equation*}
  \zbar' - \zbar = \GDs \zbar = \sum q_i'z_i - \sum q_iz_i=\sum \left(q_i'-q_i\right)z_i,
\end{equation*}
where $\GDs$ means the change caused by selection \autocite{price72fishers,ewens89an-interpretation,frank92fishers}.  We may simplify this expression by using $\GD q_i=q_i'-q_i$ for frequency changes
\begin{equation}\label{eq:charChange}
  \GDs \zbar = \sum \GD q_i z_i.
\end{equation}
This equation expresses the fundamental concept of selection \autocite{frank12naturalb}.  Frequencies change according to differences in fitness, as given by \Eq{aveEx}.  Thus, \Eq{charChange} is the change in character value caused by differences in fitness, holding constant the character values, $z_i$.  Later, we will also include the changes in character values during transmission from parent to offspring, $\GD z_i=z_i'-z_i$.  

\subsection*{Variance, covariance and regression} 

Many of the classic equations of selection are expressed in terms of variances, covariances and regressions. I show the relation between the expression for frequency changes in \Eq{charChange} and the common statistical expressions for selection.  

Combining eqns~\ref{eq:aveEx} and \ref{eq:charChange} leads to
\begin{equation*}
  \GDs \zbar = \sum \GD q_i z_i = \sum q_i\left(\frac{w_i}{\wbar} - 1\right)z_i.
\end{equation*}
On the right-hand side, move the $\wbar$ term outside
\begin{equation}\label{eq:charChange2}
  \GDs \zbar = \sum q_i\left(\frac{w_i}{\wbar} - 1\right)z_i= \sum q_i\left(w_i-\wbar\right)z_i/\wbar.
\end{equation}
The definition of the population covariance allows us to rewrite this equation. Given a population of paired values $(x_i,y_i)$, where each particular pair subscripted by $i$ occurs at frequency $q_i$, and writing $\bar{x}$ as the mean value in the population of the $x$ values, the population covariance has the general form

\begin{figure}[H]
\begin{minipage}{\hsize}
\parindent=15pt
\noterule
{\bf \noindent\BoxLabel. Interpretation of $q'$ and $z'$}
\noterule
\noindent Classical population genetics and replicator equation analyses interpret $q_i'$ in \Eq{replicator} as the frequency of type $i$ in the descendant population.  However, selection theory in its most abstract and general form requires a set mapping interpretation, in which $q_i'$ is the frequency of descendants derived from type $i$ in the ancestral population.  The set mapping interpretation arises from the Price equation \autocite{price72extension,frank95george,frank97the-price,frank98foundations}.

Similarly, $z_i'$, developed in \Eq{price} and mentioned earlier, is the average value of the property associated with $z$ among the descendants derived from ancestors with index $i$, rather than the usual interpretation of the character value of $i$ types in the descendant population.   Here, I elaborate briefly on these interpretations of $q'$ and $z'$ by adapting the presentation in \textcite{frank12naturalb}.

Let $q_i$ be the frequency of the $i$th type in the ancestral population.  The index $i$ may be used as a label for any sort of property of things in the set, such as allele, genotype, phenotype, group of individuals, and so on.  Let $q'_i$ be the frequencies in the descendant population, defined as the fraction of the descendant population that is derived from members of the ancestral population that have the label $i$.  Thus, if $i=2$ specifies a particular phenotype, then $q'_2$ is not the frequency of the phenotype $i=2$ among the descendants.  Rather, it is the fraction of the descendants derived from entities with the phenotype $i=2$ in the ancestors.  One can have partial assignments, such that a descendant entity derives from more than one ancestor, in which case each ancestor gets a fractional assignment of the descendant.  The key is that the $i$ indexing is always with respect to the properties of the ancestors, and descendant frequencies have to do with the fraction of descendants derived from particular ancestors.  

Given this particular mapping between sets, we can specify a particular definition for fitness.  Let $q'_i = q_i(w_i/\wbar)$, where $w_i$ is the fitness of the $i$th type and $\wbar = \sum q_iw_i$ is average fitness.  Here, $w_i/\wbar$ is proportional to the fraction of the descendant population that derives from type $i$ entities in the ancestors.  

Usually, we are interested in how some measurement changes or evolves between sets or over time.  Let the measurement for each $i$ be $z_i$.  The value $z$ may be the frequency of a gene, the squared deviation of some phenotypic value in relation to the mean, the value obtained by multiplying measurements of two different phenotypes of the same entity, and so on.  In other words, $z_i$ can be a measurement of any property of an entity with label, $i$.  The average property value is $\zbar=\sum q_iz_i$, where this is a population average.  

The value $z'_i$ has a peculiar definition that parallels the definition for $q'_i$.  In particular, $z'_i$ is the average measurement of the property associated with $z$ among the descendants derived from ancestors with index $i$.  The population average among descendants is $\zbar'=\sum q'_iz'_i$.

The Price equation (\Eq{price}) expresses the total change in the average property value, $\GD\zbar=\zbar'-\zbar$, in terms of these special definitions of set relations.  This way of expressing total evolutionary change and the part of total change that can be separated out as selection is very different from the usual ways of thinking about populations and evolutionary change.  The set mapping interpretation allows one to generalize equations of selection theory and  total evolutionary change to a much wider array of problems than would be possible under the common interpretations of the terms.  By following the set mapping approach, our evaluation of selection and information can be presented in a much simpler and more general way.  Note that the classic interpretations of the haploid and replicator models are special cases of the generalized set mapping expressions.    
\noterule
\end{minipage}
\end{figure}
\boxlabel{price}

\begin{figure}[H]
\begin{minipage}{\hsize}
\parindent=15pt
\noterule\vskip1.5pt
{\bf \noindent\BoxLabel. Selection and information}\vskip1.5pt
\noterule
\noindent No one seems to have provided a full development of the relations between selection and information. In many respects, R.~A.\ Fisher created the key concepts.  However, before I start listing aspects of the problem and related citations, I cannot resist quoting from \textcite[p.~96]{li08an-introduction} about the difficulties of attribution.  In discussing the name ``Kolmogorov complexity'' for the discipline of the algorithmic analysis of complexity, they note that Solomonoff published the key idea before Kolmogorov, although Kolmogorov later discovered the idea independently and developed it more deeply and thoroughly.  Ultimately, Kolmogorov got almost all the credit, perhaps because he was much more famous than Solomonoff.  Li \& Vit{\'a}nyi summarize as follows.
\begin{quote}
Associating Kolmogorov's name with the area may be viewed as an example in the sociology of science of the Matthew effect, first noted in the Gospel according to Matthew, 25: 29--30, ``For to every one who has more will be given, and he will have in abundance; but from him who has not, even what he has will be taken away.''
\end{quote}

\textcite{fisher30the-genetical} discussed the relation of his fundamental theorem of natural selection to the second law of thermodynamics, a universal law about changes in entropy.  However, Fisher never came around to an information perspective in this discussion and, perhaps for that reason, was restrained in his enthusiasm for the analogy.  Alternatively, Fisher's restraint may have had to do with the high dimensionality of the evolutionary problem \autocite{edwards00fisher}.  However, one of Fisher's great contributions in his book was his use of the average effect to reduce the dimensionality required for analyzing selection.  Although, Fisher never developed an information analysis of selection, one must remember that the modern field of information theory only began with Shannon's work on communication \autocite{shannon48a-mathematical-a,shannon48a-mathematical-b}.  The use of Fisher information outside of statistical problems developed later.

The analogy between selection and information is obvious and has been mentioned often.  However, brief mention of the analogy does not, by itself, provide any real insight about the connections between information and selection or new ways in which to understand selection.

\textcite{edwards00fisher} noted that, in the continuous-time limit, the fundamental equations of selection can be expressed in terms of Fisher information.  However, he concluded that the analogy between selection and Fisher information provides little insight.  By contrast, \textcite{frieden01population} argued that selection expressed in terms of Fisher information is indeed significant.  Although I believe Frieden et al.\ were on the right track, their particular analysis and presentation did not add much.  Fisher information is always information about an underlying scale.  Frieden et al.\ concluded that natural selection provides a measure of Fisher information about time, which I think is the wrong scale on which to interpret meaning. The present article extends the start made in \textcite{frank09natural}.    
\noterule
\end{minipage}
\end{figure}
\boxlabel{selectInfo}

\newpage

\begin{equation*}
  \sum q_i(x_i-\bar{x})y_i=\cov(x,y).
\end{equation*}
Note that the right-hand expression in \Eq{charChange2} has the form of the covariance definition, so we can write
\begin{equation}\label{eq:charCov}
  \GDs \zbar = \sum q_i\left(w_i-\wbar\right)z_i/\wbar=\cov(w,z)/\wbar,
\end{equation}
following \textcite{price70selection}. The standard definition of a regression coefficient of $y$ on $x$ is the covariance of $y$ and $x$ divided by the variance of $x$.  Thus, the regression of fitness, $w$, on character, $z$ is
\begin{equation}\label{eq:regwz}
  \Gb_{wz} = \frac{\cov(w,z)}{V_z}
\end{equation}
where $V_z$ denotes the variance of $z$. This expression implies $\cov(w,z)=\Gb_{wz}V_z$. We can also reverse the order of the regression, $\cov(w,z)=\Gb_{zw}V_w$. Thus, \Eq{charCov} is equivalently
\begin{equation}\label{eq:charReg}
  \GDs \zbar = \Gb_{wz}V_z/\wbar = \Gb_{zw}V_w/\wbar.
\end{equation}
Because $z$ can be the value of any character, we can use fitness, $w$, in place of $z$, yielding
\begin{equation}\label{eq:meanFit}
  \GDs \wbar = V_w/\wbar,
\end{equation}
where the regression has disappeared because the regression of a variable on itself is one, thus $\Gb_{ww}=1$. This expression shows that the change in mean fitness is the variance in fitness, normalized by the initial mean value.

All of these expressions assume that character values do not change between parent and offspring, $\GD z_i=0$.  As I mentioned, I will take up changes during transmission in a later section.    

\section*{Selection expressed as change in information}

This section derives a new result that connects the change in fitness caused by natural selection to the amount of information accumulated by the population.  In particular, I express the change caused by selection in terms of a classical measure of information from formal information theory.  Those readers unfamiliar with information theory will find some new expressions in this section, presented without explanation.  The following sections explain the meaning of the expressions from information theory and the connection to natural selection. (See Boxes \ref{box:selectInfo}--\ref{box:bayes} for prior work on selection and information.)  

\begin{figure}[H]
\begin{minipage}{\hsize}
\parindent=15pt
\noterule
{\bf \noindent\BoxLabel. Entropy, information and stochastic evolutionary models}
\noterule
\noindent The most interesting development of the theory arises from stochastic models of evolutionary change framed in terms of entropy and statistical mechanics.  \textcite{iwasa88free} derived a general expression for ``free fitness'' by analogy with free energy and entropy.  Iwasa showed the analogy between the continual increase of free fitness in evolutionary models and the second law of thermodynamics, by which entropy continually increases.  He also calculated the distributions in population characteristics as they change under various stochastic models of evolutionary change.  

These kinds of stochastic evolutionary models require certain assumptions in order to achieve continual increase in entropy or free fitness. There is certainly no universal law about the increase of fitness in evolution, whereas restricted notions of selection may have universal properties.  I have drawn a sharp distinction between selection and evolution in my own analyses.  The evolutionary literature does not always make that distinction so clearly.

\textcite{vladar11the-contribution} reviewed the significant advances in the use of entropy and statistical mechanics to study evolutionary dynamics, including their own contributions to the subject \autocite{barton09statistical,vladar11the-statistical}.  This work on stochastic evolutionary models may eventually converge with general studies of entropy, information and dynamics.  For example, there has been recent discussion about a maximum entropy production (MEP) principle for dynamics \autocite{dewar05maximum,kleidon10a-basic,volk10it-is-not-the-entropy}. In the MEP theory, the most likely dynamical path is associated with the greatest production of entropy.  Further, the probability distribution over dynamical paths may be a function of the relative entropy production associated with the different paths.  

One may be able to use the distribution of entropy changes over paths to calculate the stochastic evolution of populations.  Under some conditions, one may be able to specify the expected probability distribution over types when the population achieves certain kinds of equilibrium.  However, a full understanding of MEP and its limitations has yet to be achieved.  There may be some relation between dynamics analyzed in terms of Fisher information \autocite{frieden04science} and MEP. However, I do not understand the similarities and differences of those approaches.  
\noterule
\end{minipage}
\end{figure}
\boxlabel{entropy}
 
\subsection*{Change in log fitness}

Fitness captures the notion of a match between a type and the environment.   We may therefore expect that fitness is, in some way, an expression of the information in the population about the environment.  Those types with high fitness increase in frequency, increasing the fitness (information) contained in the population.  

From \Eq{replicator}, we can write the fitness of a type, $w_i$, in terms of current frequencies, $q_i$, and updated frequencies after selection, $q_i'$, as
\begin{equation*}
  w_i = \wbar\left(\qfrac\right).
\end{equation*}
Fitness depends on the ratio of frequencies, $q_i'/q_i$. Entities that depend on ratios have a natural logarithmic scaling \autocite{hand04measurement}. Therefore, we should use the logarithmic scale when analyzing fitness \autocite{wagner10the-measurement}.  It is traditional to describe the logarithm of fitness as the Malthusian expression, $m_i = \log(w_i)$, yielding
\begin{equation*}
  m_i = \log(w_i) = \log(\wbar) + \qfraclog.
\end{equation*}
Using $z\equiv m$ as our character in the selection expression of \Eq{charChange2}, we have the increase in mean log fitness by natural selection as
\begin{equation}\label{eq:changeM}
  \GDs\mbar = \sum \GD q_i \qfraclog.
\end{equation}

\subsection*{An information measure for the change in fitness}

Perhaps the most important measure of information in communication, statistics and physics is the Kullback-Leibler divergence
\begin{equation}\label{eq:KLdef}
  \KL{q'}{q} =\sum q_i'\qfraclog.
\end{equation}
This divergence has directionality from the initial population, $q$, to the updated population after selection, $q'$ (see \Boxx{info}).   Using this definition for $\D$ in the expression for the change in fitness given in \Eq{changeM}, we obtain
\begin{equation}\label{eq:changeMD}
  \GDs\mbar = \KL{q'}{q} + \KL{q}{q'}.
\end{equation}
This expression is the sum of Kullback-Leibler divergences taken in each direction between the initial population, $q$, and the updated population after selection, $q'$. In information theory, this sum is known as the Jeffreys divergence
\begin{equation}\label{eq:J}
  \J(q',q) = \KL{q'}{q} + \KL{q}{q'}.
\end{equation}
Thus, we have the simple expression for the change in mean log fitness caused by natural selection as 
\begin{equation}\label{eq:mJ}
 \GDs\mbar= \J
\end{equation}
where $\J$ is shorthand for $\J(q',q)$.  Equating this expression with \Eq{charReg}, using $m\equiv z$, we have
\begin{equation}\label{eq:Jvar}
 \J = \Gb_{wm}V_m/\wbar = \Gb_{mw}V_w/\wbar,
\end{equation}
Thus, the variance in fitness is proportional to the information divergence, $\J$.  The regression terms divided by $\wbar$ give the constants of proportionality that adjust for the different scales of measurement for fitness, $w$ or $m=\log(w)$.  This expression shows the relation between the information accumulated by natural selection, $\J$, and the traditional statistical expressions of natural selection in terms of variances and regression coefficients.

\begin{figure}[H]
\begin{minipage}{\hsize}
\parindent=15pt
\noterule
{\bf \noindent\BoxLabel. Bayesian interpretations of selection}
\noterule
\noindent Bayesian updating combines prior information with new information to improve prediction.   The Bayesian process makes an obvious analogy with selection.  The initial population encodes predictions about the fit of characters to the environment.  Selection through differential fitness provides new information.  The updated population combines the prior information in the initial population with the new information from selection to improve the fit of the new population to the environment.  I am sure this Bayesian analogy has been noted many times.  But it has never developed into a coherent framework that has contributed significantly to understanding selection.  

Part of the problem is that the analogy, as currently developed, provides little more than a match of labels between the theory of selection and Bayesian theory.  As \textcite{harper10the-replicator} shows, if one begins with the replicator equation (\Eq{replicator}), then one can label the set $\{q_i\}$ as the initial (prior) population, $\{w_i/\wbar\}$ as the new information through differential fitness, and $\{q_i'\}$ as the updated (posterior) population.   \textcite{shalizi09dynamics} presents a similar view.  The analogy provides a useful correspondence between the structure of the theories but, by itself, does not provide any truly significant insight into selection. It may be possible to develop the analogy in useful ways, a challenge that remains open.

Another Bayesian line of study analyzes how individuals adjust their characters in response to information obtained directly from the environment.  Those studies include learning, phenotypic plasticity, and various aspects of conditional development.  By one view, learning and other processes that accumulate information follow Popper's \citeyear{popper72objective} dictum that all new knowledge must ultimately derive from trial and error, in effect, from selection. 

Vast literatures discuss information theoretic and Bayesian interpretations of learning, which are beyond our scope.  In an explicitly selectionist view, \textcite{fernando12selectionist} analyze theories of neural development in relation to Bayesian updating---part of the wider field of developmental selection \autocite{frank96the-design,frank97developmental,frank97the-design}. Closer to the standard evolutionary interpretation of selection, \textcite{donaldson-matasci10the-fitness} provide an interesting discussion of information directly acquired from the environment in relation to fitness.  \textcite[Section 6.3]{frank98foundations} used a Bayesian analysis to combine selectively acquired information by the population as a prior state with new information acquired directly from the environment (learning).  
\noterule
\end{minipage}
\end{figure}
\boxlabel{bayes}
 
\section*{The encoding of information}

Before continuing to discuss the relation between selection and information, we need some additional background about the nature of information.  I first describe an example in which an observation provides information. I then discuss how to quantify the amount of information.  Finally, I analyze the amount of information in a comparison, which provides the basis for comparing the information in a population before and after selection.

\subsection*{Statistics and information}

In statistical problems, the divergence, $\D$, measures the amount of information in an observation with respect to discriminating between two distributions \autocite{kullback59information,cover91elements}.  Suppose the true underlying probability distribution is $q'$.  However, we do not know whether we are sampling from $q'$ or an alternative distribution $q$.  The different distributions may be associated with different values of a parameter, $\Gth'$ and $\Gth$. The parameter may, for example, be the mean or the variance.

When we take a sample from the true underlying distribution, $q'$, how much information do we obtain about whether the sampled distribution is $q'$ or $q$?  In the parametric case, how much information do we obtain about whether the parameter of the distribution from which we sampled is $\Gth'$ or $\Gth$?  

For each observation, with value associated to the index $i$, the relative likelihood of obtaining that observation from the true distribution, $q'$, versus the alternative distribution, $q$, is the ratio $q_i'/q_i$.  The log of the likelihood ratio is $\log(q_i'/q_i)$.  Because the true distribution is $q'$, the actual probability of observing $i$ is $q_i'$.  Thus, averaging the log-likelihood ratio over the probability of each observed $i$ value gives the average log-likelihood ratio, which is 
\begin{equation*}
  \KL{q'}{q} =\sum q_i'\qfraclog.
\end{equation*}
The divergence $\D$ is simply the average log-likelihood ratio, which means an average of the relative weight of evidence in favor of $q'$ as the true distribution compared with $q$.  The greater the ratio of likelihoods, the greater the divergence between distributions, and the greater the information in each observed value to discriminate between the distributions.

\subsection*{The scale of information}

Clearly $\D$ gives a measure of information provided by an observed value.  But what sort of scale, or units, does that measure have?  If, for example, $\D=2$, then what does the value ``two'' mean?  

The Shannon measure of information is commonly used.  That measure is related to entropy, which means randomness.  The more random something is, the less information we have about it.  For example, if a flipped coin comes up on either side with equal probability, we say that it is completely random. We also say that we have no information about which side is likely to come up.  The Shannon measure captures this duality between increasing randomness and decreasing information or, equivalently, between decreasing randomness and increasing information.

The Shannon measure is
\begin{equation}\label{eq:shannon}
  H(q) = -\sum q_i\log(q_i).
\end{equation}
We can use any base for the logarithm.  It is sometimes convenient to use base 2, in which case $H$ is the average number of bits required to encode a message.  This bit-encoding interpretation arises from the fact that 
\begin{equation*}
  -\log_2(q_i)=\log_2(1/q_i)
\end{equation*}
expresses the number of bits required to encode a probability.  For example, if $q_i$ is $1/32$, then $-\log_2(1/32)=\log_2(32)=5$ bits. A bit is the number of digits in base two required to express a number.  The number 32 in base 2 is $10000$, a bit-string with 5 digits. Each digit is a bit that takes on a value of either 0 or 1.  

To encode a probability $1/32$ requires 5 bits.  By contrast, to encode a probability of $1/2$ requires only $\log_2(2)=1$ bit.  It takes 4 bits more to encode $1/32$ compared with $1/2$.  The key idea is that a rarer event, with lower probability, $q$, provides greater surprise when the event actually occurs.  A greater surprise means a greater distinction from what was expected, a lower ability to predict, more randomness and less information.   Thus, more bits means more randomness and less information, providing a scale for measuring information in terms of bits.  

The number of bits associated with each probability concerns only that particular probability.  How should we measure the randomness and information over a set of different possible outcomes?  For a distribution, $q$, with different probabilities $q_i$ for each outcome, $i$, we must combine the randomness (bits) associated with each probability, $-\log_2(q_i)$, and the chance that the event $i$ occurs, $q_i$.  

In particular, the randomness associated with each event is the product of how often the event happens multiplied by the randomness of that event, $-q_i\log_2(q_i)$.  The total over all events is the sum given in the definition for $H(q)$ in \Eq{shannon}, which measures the total randomness over a set of events.  

To understand the notion of total randomness over a set, we can think of each $i$ as a symbol to be communicated or an event that may occur.  A message, or a set of events, has frequencies $q_i$.  In such a set, each $-\log_2(q_i)$ is the number of bits required to encode each $i$, and the event $i$ occurs with frequency $q_i$, so $-q_i\log_2(q_i)$ is the relative cost in terms of bits required to encode event $i$.  If the message, or set, is highly random, it takes a lot of bits to encode the message. High randomness corresponds to a high average level of surprise per event, which means that we have relatively little information.  

Note that information is the opposite of randomness and entropy. The measurement of information can be expressed as the negative entropy, $-H$.  

\subsection*{The information in a comparison}

The problem with $-H$ as a measure of information is that, by itself, it does not give a sense of comparison or information gain.  In the statistical example, we compared two distributions and the information gained to discriminate between those distributions provided by an observation.  In terms of selection, we will be concerned with the information gain by a population before and after evolutionary change, requiring a comparison between the initial and updated probability distributions that describe the population before and after selection.  

In a comparison, one way to measure a gain in information is by the reduction in the number of bits required to encode, or to predict, the distribution of outcomes in one population relative to another.  A reduced number of bits corresponds to reduced randomness, and reduced randomness corresponds to improved prediction and more information.  Thus, we can measure information gain by the reduction in the number of bits.

To make comparisons, we need an expanded definition of entropy
\begin{equation}\label{eq:shannon2}
  H(r,p) = -\sum r_i\log_2(p_i),
\end{equation}
where $H(r,p)$ is the entropy in the probability distribution $r$ when encoded by the associated probabilities $p$.  This expression may be interpreted by thinking of the different $i$ values as symbols in an alphabet, the $r_i$ as the frequency of the symbols in a message, and the $p_i$ as the frequencies used to determine the encoding of the symbols $i$.  Then $H(r,p)$ is the average number of bits required to encode a message $r$ in a code based on $p$. 

To compare populations, suppose an updated population has probabilities of types (events) $q_i'$, and entropy $H(q',q') = H(q')$.  By contrast, the entropy of the new population, when using the encoding of the old population, $q$, before new information was acquired, is $H(q',q)$, which is the randomness in the new population when encoded by the old frequencies.  

In the updated population, the change in information obtained from the updated encoding is the average number of bits to encode $q'$ based on the new frequencies, $H(q',q')$, minus the average number of bits to encode $q'$ based on the old frequencies, $H(q',q)$, which is
\begin{align}
  -\left(H(q',q') - H(q',q)\right) &= \sum q_i'\log_2(q_i') - \sum q_i'\log_2(q_i) \notag\\
                                                 & =\sum q_i'\log_2\left(\frac{q_i'}{q_i}\right) \notag\\
                                                 &=\KL{q'}{q}, \label{eq:bitEncode}
\end{align}
where the initial minus sign is used to express negative entropy, which is information.  The term $\log_2(q_i'/q_i)$ is the number of extra bits to encode $q_i'$ given a prior assumption that event $i$ happens with probability $q_i$.  The expression $\D$ measures the average number of extra bits needed when encoding the  new population by the old frequencies rather than with the new, updated frequencies.  Thus, $\D$ is the average gain in information in a population update when measured in terms of number of bits. A value of $\D=2$ means that an efficiency gain of two bits has been achieved by the extra information provided. Alternatively, we may say that the new information enhances predictability, such that the remaining randomness, or unpredictability, has been reduced by two bits.

\section*{Selection and the meaning of information}

The encoding interpretation of information is well known and widely accepted \autocite{kullback59information,cover91elements}.  By contrast, a formal interpretation of natural selection in terms of information has never been developed in a simple, clear, and widely agreed manner.  Here, I give my interpretation of natural selection and information.

\subsection*{Why $\J$ rather than $\D$?}

To analyze meaning of information with regard to natural selection, we must begin with the fundamental expression of selection in terms of information divergence given in \Eq{mJ} as $\GDs\mbar=\J$. That expression states that the change in mean log fitness is the Jeffreys divergence, $\J$.  Recall the definition of $\J$ from \Eq{J} as
\begin{equation*}
  \J(q',q) = \KL{q'}{q} + \KL{q}{q'}.
\end{equation*}
In most statistical and physical applications, measures of divergence and information typically use $\D$ \autocite{cover91elements}.  For example, Bayesian updating can often be expressed in terms of a prior distribution, $q$, an updated distribution based on new data, $q'$, and the divergence of the updated distribution from the prior, $\KL{q'}{q}$.  In the Bayesian expression, $\D$ describes the gain in information measured in terms of bits and interpreted with regard to the efficiency of encoding information or, equivalently, the reduced randomness and increased predictability of outcomes. 

The measure $\D$ is asymmetric, because $\KL{q'}{q}\ne\KL{q}{q'}$.  By contrast, $\J$ is symmetric, because it is the sum of the divergence in each direction.  The symmetry in the selection equation arises because, from \Eq{changeM}, we have
\begin{align}
  \GDs\mbar &= \sum \GD q_i \qfraclog \notag\\
                   &= \sum \GD q_i \left[\log(q_i')-\log(q_i) \right] \notag\\
                   &= \sum \GD q_i\left[\dlog \right]. \label{eq:dd}
\end{align}
If we switch $q_i'$ and $q_i$, then $\GD q_i$ changes sign, and $\dlog$ also changes sign.  The two sign changes cancel.  Thus, we obtain the same information gain when selection moves a population as $q\rightarrow q'$ or in the reverse direction as $q'\rightarrow q$.  

\subsection*{Fitness in terms of encoded information}

The information expression for fitness in \Eq{dd} is in terms of $\log(q_i'/q_i)$. Thus, the information gain continues to be about efficiency of encoding or, equivalently, the reduced randomness and increased predictability of outcomes.  We could, for example, think of an increase in mean log fitness as an increase in the population's prediction of, or match to, the state of nature---the fit of the population to the environmental challenge.  

This interpretation of fitness in terms of encoding is universal, in the sense that the particular environmental challenges and the particular meaning of the gain in fitness with respect to particular characters do not enter into the expressions.  The universal expression of fitness and selection in terms of probabilities and encoding yields the match between changes in mean log fitness and changes in the classical expressions of information.

\subsection*{Encoding versus meaning}

The great power and universality of the classic theory of information arises because it does not depend on meaning.  Information is formulated strictly in terms of encoding, bits, randomness, and predictability, independently of what is being encoded or predicted.  Fitness obtains the same universality, because fitness uses the same expressions of relative frequency as the classic information measures.  That universality for fitness makes sense, because fitness is a general expression for the way in which populations accumulate information, independent of the characters and environmental challenges that distinguish particular cases.

Although it is certainly beneficial to have a universal expression of fitness in terms of information, we pay for that universality by the limited scope of fitness expressed only in terms of encoding.  Information is about predictability, and predictability is always predictability about something.  Natural selection must, in some way, be about the increased information with respect to the environmental challenges that shape success.  How can we bring this particular meaning of the information about environmental challenges into the formulation of fitness?

There is perhaps no universal way to express meaning with respect to information.  That may be why the encoding interpretation has been so valuable.  The following sections explore two related ways in which to bring meaning into the information interpretation of fitness.  The next section develops the notion of Fisher information.  Later sections present the idea of a coordinate system for information and evolutionary change---a connection between the Price equation and information.

\section*{Natural selection and Fisher information}

\begin{quote}
Shannon information is not really information as such, but rather the capacity to transmit information, whereas Fisher information is truly a measure of informativeness about something specific, the value of a parameter. Shannon's refers to the medium, Fisher's to the message \autocite[p.~6]{edwards00fisher}.
\end{quote}

\noindent We have been working on the scale of encoded information. That scale depends only on probability distributions, without any explicit connection to what sort of events or meaning attach to the probabilities.  Units of encoded information can be measured in terms of bits. \autocite[The following extends][]{frank09natural}.

One way to interpret meaning is to change the scale.  Suppose we could relate bits of encoded information to a new scale on which we interpret meaning.  To relate the change in information to the change in meaning, we could evaluate
\begin{equation}\label{eq:meaning}
  \GD\mathrm{information} = \left(\frac{\GD\mathrm{information}}
  				{\GD\mathrm{meaning}}\right) \GD\mathrm{meaning}.
\end{equation}
The relation is trivial when expressed in this way. However, we can see that the ratio of change in information to change in meaning provides the translation between the two scales.   

To make this expression for the relations between the scales useful, we must connect each of the terms to our prior discussion of information and to a new way of describing meaning.  That connection leads us to expressions of natural selection in terms of the fit of characters to the environment, rather than the efficiency of encoding information in terms of bits.  

Up to this point, I have been writing $q_i$ or $q_i'$ for the probability of event $i$, whatever sort of event or characteristic $i$ may be.   The probability distribution is the set of $q_i$ values over the range of possible characters, each possible character associated with a label $i$.  In this formulation, one can think of the probability distributions as interpreted nonparametrically, in the sense that we work directly with the actual distribution of probabilities without reference to any underlying parameters or causes.  

Now suppose we associate a set of values, $\Gth$, with each probability distribution \autocite{amari00methods}.  We could think of $\Gth$ as a parameter, for example, the mean of the distribution.   Or we could think of $\Gth$ as the predictions about the environment associated with a probability distribution.  The predictions might be expressed as characters. The quality of the predictions could be associated with fitness.  

For now, we take $\Gth$ in the general sense of some values associated with a distribution. To express the association, we expand our notation for probabilities to write $q_i|\Gth$, the probability of event $i$ given the associated value $\Gth$.  An updated population may have a new associated value, $\Gth'$, such as a new mean or a new prediction about the environment, so we write $q_i'|\Gth'$.  The change in probability is now expressed as 
\begin{equation*}
\GD q_i|\Gth = q_i'|\Gth' - q_i|\Gth.
\end{equation*}
To express the scaling of probability changes relative to changes on the new $\Gth$ scale, we can divide both sides by the change on the $\Gth$ scale, yielding
\begin{equation*}
\frac{\GD q_i|\Gth}{\GD\Gth} =\frac{q_i'|\Gth' - q_i|\Gth}{\Gth'-\Gth}.
\end{equation*}
This expression gives us a way to match changes on the scale of meaning, $\Gth$, to changes on the scale of probability and encoded information, $q$. 

We can now follow \Eq{meaning} to express the change in information as the change on the scale of meaning multiplied by the change of information scaled relative to the change in meaning.  To develop this expression, we must continue to match our previous work on information and selection to the new notation in relation to meaning.  

The log-likelihood ratio, $\log(q_i'/q_i)$, can be written as $\log(q_i') - \log(q_i)$, which may be abbreviated as $\GD\log(q_i)$, as in \Eq{dd}.  This difference of logarithms expresses the change in the number of bits required to encode the probabilities associated with $i$ (as described below \Eq{bitEncode}).  If we now express probabilities in relation to $\Gth$, as $q|\Gth$, and divide by $\GD\Gth$, we obtain the change in the number of bits in relation to the change on our scale of meaning
\begin{equation*}
 \frac{\log(q_i'|\Gth')-\log(q_i|\Gth)}{\Gth'-\Gth} = \frac{\GD\log(q_i|\Gth)}{\GD\Gth}.
\end{equation*}
We can now put the pieces together by relating these new expressions with the expression in \Eq{dd} for the change in mean log fitness, yielding a form equivalent to the intuitive description in \Eq{meaning} as
\begin{equation}\label{eq:mF1}
  \GDs\mbar = \frac{\J(\Gth)}{\GD\Gth^2}\GD\Gth^2,
\end{equation}
in which I write $\GD\Gth^2=(\GD\Gth)^2$ for the square of the change in the parameter, and the term $\J(\Gth)$ is the Jeffreys divergence, which is now a function of the scale of  meaning, $\Gth$, and is written as
\begin{equation}\label{eq:Jparam}
  \J(\Gth) = \sum \left(\GD q_i|\Gth\right)\left[\GD\log(q_i|\Gth)\right].
\end{equation}
These expressions simply repeat our prior derivation of $\GDs\mbar = \J$, but with explicit consideration of $\Gth$.

As the changes become small, $\GD\Gth\rightarrow 0$, the Jeffreys divergence, $\J(\Gth)$, divided by the squared change in scale, $\GD\Gth^2$, converges to the important quantity in statistical theory known as Fisher information, $F(\Gth)$, which we write as
\begin{equation*}
  \frac{\J(\Gth)}{\GD\Gth^2} \rightarrow F(\Gth),
\end{equation*}
as shown in Appendix A.  Thus, for small changes on the scale of meaning, $\GD\Gth\rightarrow 0$, we may write the change in average log fitness as
\begin{equation}\label{eq:infoSelect}
  \GDs\mbar = F(\Gth)\GD\Gth^2.
\end{equation}
This derivation provides a more general way to arrive at my earlier statement that changes in mean fitness are proportional to Fisher information \autocite{frank09natural}.  Fisher information is the information in an observation about a parameter, or a set of parameters.  In our case, $\Gth$ represents the parameters, which is our scale of meaning.  

One can also think of Fisher information as the Jeffreys divergence between populations, $\J(\Gth)$, relative to the squared divergence on the scale of meaning, $\GD\Gth^2$.  Thus, Fisher information is the sensitivity of change in the encoded information in populations, $\J(\Gth)$, relative to change on the parametric scale of meaning.  The greater the sensitivity, the more information in an observation with respect to the divergence between populations on the underlying parametric scale. See Appendix B for ways in which Fisher information has been used in previous models of selection.

\section*{Parametric coordinates for selection and information}

The change in mean log fitness measures the amount of information that the population accumulates by selection.  Because fitness describes changes in relative frequencies, fitness concerns encoding of information, which can be measured in numbers of bits.  

The previous section showed how to convert from bits to an alternative scaling of information in terms of $\Gth$.  We may interpret the parameters $\Gth$ as a scale that has meaning with respect to the fit of the population's characteristics to the environment.  This section further analyzes the notion of parametric coordinates for selection and information, followed by an example.

\subsection*{Parametric coordinates and Fisher information}

From \Eq{mF1}, the key result for the change in mean log fitness in terms of a parametric scale can be rewritten as
\begin{equation}\label{eq:fisherApprox}
  \frac{\GDs\mbar}{\GD\Gth^2} = \frac{\J(\Gth)}{\GD\Gth^2}\rightarrow F(\Gth).
\end{equation}
Change in mean log fitness is the amount of information gained by selection.  The ratio $\GDs\mbar/\GD\Gth^2$ is the change in information per unit change in squared distance on the parametric scale.  Because we consider the parametric scale as the scale of meaning, this ratio is the change in information relative to the change in squared distance on the scale of meaning \autocite{amari00methods}.  The arrow on the right-hand side states that the relative change in information per unit of squared parametric distance is the Fisher information in an observation about the parameter, $\Gth$.

The interpretation of ``observation'' with respect to natural selection is interesting. Each interaction of an individual  with the environment leads to a realized fitness.  That realized individual fitness is an observation, by the population, of the fit between certain characteristics and the environment.  For a particular type, $i$, the average information in each observed individual fitness is $\log(q_i'/q_i)=\GD\log(q_i|\Gth)$. Thus, the ratio $\GD\log(q_i|\Gth)/\GD\Gth$ is the change, or sensitivity, of information in an observation relative to a change in $\Gth$.  To get the average over all types, $i$, we weight this information per type by $q_i|\Gth$. To analyze selection, we need the change in frequencies, or sensitivity of those changes, relative to changes in $\Gth$, which is $\GD q_i|\Gth/\GD\Gth$.  Combining these terms yields $\J(\Gth)/\GD\Gth^2 \rightarrow F(\Gth)$.  

\subsection*{Change in the mean or variance of a character}

A few examples clarify the abstract expressions for information.  To keep things simple, I assume small changes so that we can use the Fisher information simplification in \Eq{fisherApprox}.  With larger changes, we could make exact calculations using $\J(\Gth)$ instead of Fisher information.

\subsubsection*{Change in the mean of a normal distribution under directional selection}

Suppose the character values in a population, $z_i$, follow a normal distribution with mean, $\Gm$, and variance, $v$.  An observation from that population provides information about the mean of the population.  It is well known that an observation from a normal population provides Fisher information about the mean of $F(\Gm)=1/v$.  The more variable the population, the larger $v$, and the less information in an observation about the average value.  Put another way, the precision in measurement is proportional to $1/v$.  More variable populations yield less precise measurements, and thus less information per observation about the average value.  

We interpret natural selection as obtaining information through the observed fitnesses associated with character values.  Suppose that the population retains a normal shape and a fixed variance before and after selection, and changes only in its mean value. Then the change in the mean, $\GD\Gm$, is sufficient to describe the effects of selection. From \Eq{infoSelect}, the increase in information by natural selection is
\begin{equation*}
  \GDs\mbar = F(\Gm)\GD\Gm^2 = \frac{\GD\Gm^2}{v}.
\end{equation*}
This expression provides the relation between the change in information, $\GDs\mbar$, which is a universal abstract quantity about encoding, and the scaling of the character that gives meaning for this particular case, $\GD\Gm^2/v$.  

\subsubsection*{Change in the variance of a normal distribution under stabilizing selection}

The previous example described directional selection on the average trait value, holding the variance constant.  This section considers stabilizing selection. In this case, the population begins with its center at the optimum. Selection reduces the variance, but leaves the mean unchanged. For a normal distribution, the Fisher information in an observation about the variance, $v$, is $1/2v^2$.  Thus, 
\begin{equation*}
  \GDs\mbar = F(v)\GD v^2 = \frac{\GD v^2}{2v^2},
\end{equation*}
which is the gain in information when stabilizing selection reduces the variance of a normally distributed character.

\subsubsection*{Change in the mean of an exponential distribution}

Suppose the character follows an exponential distribution before and after selection.  An observation from an exponential population provides Fisher information of $1/v$ about the mean, $\Gm$.  The variance of an exponential distribution is $v=\Gm^2$. The change in information by selection is
\begin{equation*}
  \GDs\mbar = F(\Gm)\GD\Gm^2 = \frac{\GD\Gm^2}{v},
\end{equation*}
which matches the case of the normal distribution.  However, the variance of the exponential distribution changes with the mean. By contrast, the normal distribution has a separate parameter for the variance, which we held constant by assumption.

\subsubsection*{Change in allele frequency}

Suppose $q_1=p$ is the frequency of a particular allele, and $q_0=1-p$ is the frequency of the alternative allele.  The distribution of allele frequencies is binomial with a single observation.  The mean allelic value is $\Gm=p$, and the variance is $v=p(1-p)$.  The Fisher information in an observation about the mean of a binomial population is $1/v$.  The change in information by selection is
\begin{equation*}
  \GDs\mbar = F(\Gm)\GD\Gm^2 = \frac{\GD\Gm^2}{v}.
\end{equation*}
Using $p$ for gene frequency to match the familiar notation of population genetics 
\begin{equation*}
  \GDs\mbar = F(p)\GD p^2 = \frac{\GD p^2}{p(1-p)},
\end{equation*}
which holds when $\GD\Gm=\GD p$ is small.  For larger changes, we can obtain an exact expression by using the Jeffreys divergence rather than the Fisher information, as in \Eq{fisherApprox}.

\section*{Character coordinates and selection}

The previous section assumed that the parameters, $\Gth$, summarize all differences in the frequency distributions before and after selection.  We can think of $\Gth$ as defining the coordinate system for evolutionary change.  The reduction of frequencies to a parametric description, such as the mean of the distribution, typically requires character values to be associated with the $i$ values.  By convention, we use $z_i$ for character values.  Thus, if changes in the mean are sufficient to describe changes in the probability distribution of characters in the population before and after selection, then $\Gm=\zbar=\sum q_iz_i$ is a reduction of the full distribution of character values to a single parametric dimension.  

\subsection*{Parametric character coordinates}

Let us review the use of parametric coordinates before discussing nonparametric coordinates.  In a parametric example, suppose that frequencies before and after selection are normally distributed, with parameters $(\Gm,v)$ for the mean and the variance.  Selection moves the population from the initial location, defined by the parameters $(\Gm,v)$, to the location after selection, $(\Gm',v')$.  The two parametric dimensions provide a complete description of change by selection.  If we hold one parameter constant, such as the variance, and only allow the mean to change, then change in the single parametric dimension from $\Gm$ to $\Gm'$ fully describes the population before and after selection.

Parametric expressions describe the total change in information by
\begin{equation*}
	\GDs\mbar =\frac{\GD \J}{\GD\Gth^2} \GD\Gth^2\rightarrow F(\Gth)\GD\Gth^2.
\end{equation*}
For example, let the parameter be the mean, $\Gth=\Gm$. The term $\J(\Gm)/\GD\Gm^2\rightarrow F(\Gm)$ reduces the change in the average information per observation to the single dimension of $\Gm$.  If we multiply the information per observation by the distance moved in the parametric dimension, $\GD\Gm^2$, we obtain the total change in information.  Thus, the calculation for the change in information is done along the single parametric dimension of $\Gm$.  

The parametric dimension of $\Gm$ can be thought of as the coordinate system in which we evaluate change by selection.  Each change in position along the coordinate of $\Gm$ corresponds to changes by selection, because $\Gm$ is a sufficient description for the full frequency distribution of character values.  In general, when we can reduce the description of frequency distributions to a sufficient set of parameters, $\Gth$, then those parameters form the coordinates in which we evaluate changes by selection.

\subsection*{Nonparametric character coordinates}

We can think of our fundamental expression for selection
\begin{equation*}
	\GDs\zbar =  \sum \GD q_i z_i
\end{equation*}
as a nonparametric expression. Each term includes the actual frequencies in the population.  The calculation is done over the full dimensionality of the frequency distribution.

The character values, $\{z_i\}=z_1,z_2,\ldots$, form a nonparametric coordinate system.  For the population frequencies, $\{q_i\}$, the point $\{q_iz_i\}$ locates the population before selection, and the point $\{q_i'z_i\}$ locates the population after selection. The movement of the population caused by selection is given by $\{\GD q_iz_i\}$.  

The expression for the total change in information caused by selection is
\begin{equation*}
	\GDs\mbar = \J = \sum \GD q_i \GD\log(q_i) = \sum \GD q_i \qfraclog.
\end{equation*}
Each frequency change, $\GD q_i$, associates with the character $z_i=\GD\log(q_i)$, the change in information for the $i$th type.  This is a nonparametric expression, because the calculation is done over the full frequency distribution.  

\subsection*{Character coordinates and information}

The character values provide the coordinates of meaning in an analysis of selection.  We can derive the relations between information and the coordinates of meaning by using the results of eqns \ref{eq:charReg} and \ref{eq:meanFit}.  From those equations, we obtain the relation between the change given the coordinates of meaning, $\GDs\zbar$, and the change given the coordinates of information, $\GDs\mbar$, as
\begin{equation}\label{eq:coordChange}
  \GDs\zbar = \left(\frac{\Gb_{zw}}{\Gb_{mw}}\right)\GDs\mbar.
\end{equation}
The term $\Gb_{zw}$ is the regression coefficient of the character values, $z$, on the fitnesses, $w$. The term $\Gb_{mw}$ is the regression coefficient of the log fitnesses, $m$, on the fitnesses, $w$.  These regressions provide an exact expression for changing the coordinates from information, $\GDs\mbar$, to characters, $\GDs\zbar$.  When the magnitudes of the changes are small, $w\rightarrow m+1$, thus
\begin{equation}\label{eq:coordApprox}
  \GDs\zbar \rightarrow \Gb_{zm}\GDs\mbar.
\end{equation}
To repeat, it is important to recognize a regression coefficient as an exact expression for the change in scale associated with a change in coordinates.  The regression is sufficient when evaluating the consequences for a change in coordinates with respect to a change in mean value.  

The underlying values, $z_i$, may themselves be nonlinear functions of other values \autocite{frank12naturalb}. For example, $z_i$ could be the product of different character values measured on each individual, or the square of some underlying character.  What matters is that we average over the $z_i$ values to get $\GDs\zbar$.  

\section*{Character coordinates and total evolutionary change}

The previous analyses have focused on the selection part of total evolutionary change.  I defined selection as the change caused by frequency differences
\begin{equation*}
  \GDs\zbar = \sum \GD q_i z_i.
\end{equation*}
The subscript $s$ emphasizes that this expression is the partial change caused by selection \autocite{price72fishers,ewens89an-interpretation,frank92fishers}.  

\subsection*{Total change in characters}

The partial change arises by holding constant the character values, such that $\GD z_i=z_i'-z_i=0$.  This assumption fixes the coordinates, $z_i$, and evaluates the meaning of changing frequencies in the context of that fixed set of coordinates.  

If the coordinates that give meaning also change, $\GD z_i\ne0$, then we must account for that change in coordinates with respect to the total evolutionary change.  In particular, the total change is the sum of the change, $\GDs$, caused by selection through varying frequencies, $q$, holding constant the coordinates, $z$, plus the change in coordinates, $\GDc$, holding constant the new frequencies in the updated population, $q'$.  We write the total change as
\begin{align}
  \GD\zbar &=\GDs\zbar + \GDc\zbar \notag\\
                  &=\sum \GD q_i z_i + \sum q_i'\GD z_i. \label{eq:price}
\end{align}
This expression is a form of the Price equation.  I devoted the prior article to a full discussion of this equation \autocite{frank12naturalb}.  Here, I focus only on those aspects that concern information. In particular, I emphasize the interpretation of $z$ as a coordinate system that gives meaning to the information basis of natural selection.

\subsection*{Total change in information}

The total evolutionary change in \Eq{price} can be used to evaluate information.  Let $z=m$, where the log fitness, $m$, provides a measure of the information accumulated by a population .  Thus,
\begin{equation}\label{eq:totalI}
  \GD \mbar =\GDs \mbar + \GDc \mbar.
\end{equation}
From \Eq{mJ}, the selection component of change is $\GDs \mbar=\J$.  In general, no simplified reduction or particular interpretation is possible for the change in coordinates, $\GDc \mbar$.  That change in coordinates arises from any environmental or extrinsic factors that may change, altering the fit of the characters to the environment.  The changes in the frequencies themselves can be an ``environmental'' change that alters fitnesses \autocite{price72fishers,ewens89an-interpretation,frank92fishers}. Thus, no general expression for total evolution change in fitness is possible other than 
\begin{equation*}
  \GD \mbar = \J + \GDc \mbar.
\end{equation*}
One can, of course, analyze particular models such as mutation-selection balance.  Mutation decays information through changes in fitness that are, on average, negative, causing a loss of information through the term $\GDc \mbar=\sum q_i'\GD m_i$.  The particular loss of information through $\GDc \mbar$ depends on the specific assumptions. By contrast, the gain in information through selection is always $\GDs \mbar=\J$.

\subsection*{Equilibrium balance between information gain and loss}

Many processes lead to an equilibrium balance between gain of information by selection and decay of information by an opposing force \autocite{frank12natural}.  Mutation-selection balance is one example.  Frequency-dependent selection is another, in which the gain in information by selection is balanced by the decay of information (fitness) caused by frequency changes. For example, in the evolution of sex ratios, making more daughters may be favored by selection. But as the number of daughters increases by selection, the advantage of making extra daughters decays.  

Although we cannot, in general, specify the change in the coordinate term, $\GDc \mbar$, we can express the equilibrium condition, $\GD\mbar=0$.  Under a balance between information gain by selection and information decay by change in coordinates, 
\begin{equation*}
  \J=-\GDc\mbar. 
\end{equation*}
It is sometimes possible to analyze particular problems by using that universal expression for the balance of forces \autocite{frank90the-distribution,frank95george}.

\subsection*{Evolution of the coordinate system}

In the previous sections, I have fixed the particular dimensions that define the coordinate system.  Although the coordinates may change, $\GD z_i$, each dimension $i$ remained.  From a broader perspective, the evolution of the various dimensions in the coordinate system itself is perhaps among the most interesting evolutionary problems. One aspect concerns the origin of new characters \autocite{west-eberhard03developmental}. More generally, one may consider the evolution of the optimal set of characters with respect to the capture of information.  

There is an interesting literature in engineering about optimal design of sensors with respect to capturing information.  That literature sometimes uses Fisher information as the optimality criterion with respect to design \autocite{borguet08the-fisher}.  Application of that design perspective with regard to information may provide insight into biological problems.  For example, multiple cellular receptors may respond to the same sort of information, such as the concentration of a hormone. But those receptors may be tuned differently with regard to sensitivity to signals.  A related idea concerns the common tradeoff between informativeness and simplicity in classification \autocite{kemp12kinship}.  

A second aspect of coordinates concerns the parametric reduction of the full nonparametric distribution of characters.  Reducing the full distribution to the mean is an extreme reduction, and probably not justified in general.  However, there often may be some suitable reduction of dimensionality to a sufficient set of parameters with respect to the acquisition of information \autocite{carter09fine:,goh11a-nonparametric}.  That sufficient set defines the coordinates of information and meaning followed by an evolving population.  It may be that an improved parametric representation of information in the environment by a set of characters enhances fitness.  Thus, it may be the parametric representation itself that is under the strongest selection or, at least, a particularly interesting form of selection.

\section*{Discussion}

The fundamental equations of selection are often written in the statistical terms of variances, covariances, and regressions.  I have argued that one obtains a deeper understanding of selection if one learns to read the fundamental equations in terms of information.  Here, I review my argument by listing the key steps derived in previous sections. I start with the classic statistical equations of selection. I then show the connection of those statistical expressions of selection to expressions for the information that populations accumulate about the fit of characters to the environment.

\subsection*{Statistical expressions of selection}

To understand where the classic statistical expressions of selection come from and what they mean, let us start with the basic equation for evolutionary change by natural selection
\begin{equation*}
  \GDs\zbar = \sum \GD q_i z_i
\end{equation*}
given in \Eq{charChange}. Here, $\GDs\zbar$ is the change caused by selection in the average value of a character, $\zbar$.  This expression applies generally to selection of any value.  For example, $z$ could be gene frequency, leading to population genetics expressions, or $z$ could be a quantitative trait such as weight, or $z$ could be a nonlinear function of several characters.  The $\GD q_i$ terms are the changes caused by selection in the frequency of the $i$th character value, $z_i$.  Total selection is the total change in frequencies, with each change caused by selection, $\GD q_i$, weighted by its associated character value, $z_i$. 

I showed that one can rewrite the association between the change caused by selection and the character value as
\begin{equation}\label{eq:discSel}
  \sum \GD q_i z_i= \cov(w,z)/\wbar,
\end{equation}
a form known as the Price equation and also related to Robertson's secondary theorem of natural selection \autocite{frank12naturalb}.  This form provides the foundation for quantitative genetics theory, and also arises in standard models of population genetics.  The definition of covariance allows us to rewrite the covariance  as the product of a regression coefficient and a variance term 
\begin{equation}\label{eq:discReg}
  \GDs\zbar = \cov(w,z)/\wbar =  \Gb_{zw}V_w/\wbar,
\end{equation}
where $\Gb_{zw}$ is the regression of character value, $z$, on fitness, $w$, and $V_w$, is the variance in fitness.  These sorts of regression and variance terms arise repeatedly in the fundamental equations of selection. 

One can easily understand why selection depends on an association between fitness, $w$, and character value, $z$.  Those character values associated with higher fitness will increase, whereas those character values associated with lower fitness will decrease.  But why should the expression for selection be exactly the covariance, or the regression multiplied by the variance, which capture only the linear component of association?  The reason is that $\GDs\zbar$ describes selection by a change in average values.  To calculate a change in the average, we need only the linear component of association between character and fitness.  

These statistical expressions of selection in terms of covariances, variances, and regressions have been very useful throughout the history of evolutionary theory.  However, these expressions give no sense of what selection means.  To say that selection is the covariance of fitness and character value is simply to express an algebraic relation.  That algebraic relation is very useful, but it does not give a sense of what selection is actually doing with regard to adaptation or how selection relates to processes in other fields of study.  The statistical expressions do not tell us how to read the fundamental equations of selection with regard to the meaning of the underlying process.

\subsection*{Selection in terms of information}

In this article, I argued that selection causes populations to accumulate information about the fit of characters to the environment.  I gave a precise definition of ``information.''  That definition of information with respect to selection matches exactly the classic usage of information and entropy from the fundamental theories of physics, statistics, and communication.  By showing the exact relations between selection and information, I tied the theory of natural selection to the broader conceptual framing of problems at the foundation of many key scientific disciplines.  

I will not repeat the whole argument here.  Instead, I list a few steps to emphasize the essential points.  To understand the information associated with selection and fitness, we must analyze fitness on a logarithmic scale
\begin{equation*}
  m_i = \log(w_i) = \log(\wbar) + \qfraclog.
\end{equation*}
The logarithmic scale compares relative magnitudes.  We need relative magnitudes because there is no meaning in the number of babies or the number of copies produced with regard to whether a type, $i$, is increasing or decreasing in the population.  We need to know the relative success. The logarithmic scale is the natural scale of relative magnitudes.

Using log fitness, $m$, as the character value of interest in \Eq{discSel}, we obtain 
\begin{equation*}
  \GDs\mbar = \sum\GD q_i m_i = \sum \GD q_i \qfraclog.
\end{equation*}
We recognize the fundamental expression for the change in information given by the Kullback-Leibler divergence, or relative entropy, as
\begin{equation*}
  \KL{q'}{q} =\sum q_i'\qfraclog.
\end{equation*}
Using this definition for change in information, $\D$, we can express the change in mean log fitness caused by selection as
\begin{equation*}
  \GDs\mbar = \KL{q'}{q} + \KL{q}{q'}.
\end{equation*}
This sum of the changes in information in each direction is known as the Jeffreys divergence, $\J$. Thus, we can write the fundamental expression for the accumulation in information by natural selection as
\begin{equation*}
  \GDs\mbar = \J.
\end{equation*}
Because $z$ in \Eq{discReg} is just a placeholder for any character, we can use $m$ in place of $z$ in that equation, yielding
\begin{equation*}
  \GDs\mbar = \Gb_{mw}V_w/\wbar.
\end{equation*}
Thus, the information accumulated by natural selection is equivalently expressed in terms of the regression coefficient and variance
\begin{equation}\label{eq:discEquiv}
  \J = \Gb_{mw}V_w/\wbar.
\end{equation}
The value of $\J$ is the gain in information.  The variance in fitness, $V_w$, is therefore a measure of the separation between the initial population and the population after selection, when the separation between populations is expressed on a scale of information.  The regression divided by the mean fitness, $\Gb_{mw}/\wbar$, is a scaling factor that translates the measure of information in $V_w$ to the scale of log fitness, $m$.  That scaling change is required because log fitness is the proper measure of information in expressions of selection.

Eqn \ref{eq:discEquiv} shows the equivalence between the expression of information gain and the expression of it terms of statistical quantities. There is nothing in the mathematics to favor either an information interpretation or a statistical interpretation.   

I have argued that, when reading the fundamental equations of selection for meaning, we should prefer the information interpretation.  The information perspective makes sense intuitively.  Selection is the process by which populations accumulate information about the environment.  

\section*{Acknowledgments}

My research is supported by National Science Foundation grant EF-0822399.  

\bibliography{main}

\newpage

\section*{Appendix A: Fisher information as the limiting form of the Jeffreys divergence}

A large family of divergence measures converges to Fisher information in the limit of small changes \autocite{amari00methods,amari10information,cichocki10families,cichocki11generalized}.  In this appendix, I show that the limit of the Jeffreys divergence is the Fisher information multiplied by a scaling factor for parametric distance.  

I also show that the chi-square divergence becomes the Fisher information metric in the limit of small changes.  The different forms of divergence can be confusing if one does not realize that all of the different divergence measures in the Fisher family are equivalent in the limit, but differ when changes are not small.

My main point is that the Jeffreys divergence holds the unique position as the only correct divergence measure for models of selection. It is the only measure that is correct both for large changes and, in the limit, for small changes.  As far as I know, my derivation in this article of the Jeffreys divergence in relation to selection has not been shown previously.  The clear relation of the Jeffreys divergence to changes in information is essential to make the proper connection between selection and information.

\subsection*{Limiting form of Jeffreys divergence}

I show $\J(\Gth) \rightarrow F(\Gth)\GD\Gth^2$ as the distance in the parametric coordinates $\GD\Gth^2 \rightarrow 0$.  Notationally, $\GD\Gth^2\equiv(\GD\Gth)^2$. Using the standard differential notation for small differences, we write $\GD\Gth^2\rightarrow\dd\Gth^2$. Thus, I show $\J(\Gth) \rightarrow F(\Gth)\dd\Gth^2$.  

I use the vector $\Gth$ as parametric coordinates for probability distributions, following standard analysis in information geometry \autocite{amari00methods}.  For simplicity, I usually treat the parametric vector as a single dimension.  The extension to multiple dimensions is standard.  

The Jeffreys divergence  in parametric form, from \Eq{Jparam}, is
\begin{equation*}
  \J(\Gth) = \sum \left(\GD q_i|\Gth\right)\left[\GD\log(q_i|\Gth)\right].
\end{equation*}
As the changes become small, $\GD q_i|\Gth=q_i'|\Gth'-q_i|\Gth\rightarrow 0$ and $\GD\Gth=\Gth'-\Gth\rightarrow 0$, we write
\begin{align*}
  \GD q_i|\Gth\rightarrow &\,\dd q_i|\Gth\\
                                       &= \left(\frac{\dd q_i|\Gth}{\dd\Gth}\right)\dd\Gth \\
                                       &= \qdot_i\dd\Gth,
\end{align*}
where $\qdot_i$ is the derivative of $q_i|\Gth$ with respect to $\Gth$.  Next,
\begin{align*}
  \GD\log(q_i|\Gth)\rightarrow &\,\dd\log(q_i|\Gth)\\
                                       &= \left(\frac{\dd\log(q_i|\Gth)}{\dd\Gth}\right)\dd\Gth \\
                                       &= \left(\frac{\qdot_i}{q_i}\right)\dd\Gth,
\end{align*}
where, to make the notation more concise, I use $q_i\equiv q_i|\Gth$. Thus
\begin{equation*}
   \J(\Gth) \rightarrow \sum \left(\frac{\qdot^2_i}{q_i}\right)\dd\Gth^2.
\end{equation*}
Below, I show that $\sum \qdot^2_i/q_i$ is Fisher information, $F(\Gth)$. Thus, $\J(\Gth)\rightarrow F(\Gth)\dd\Gth^2$. 

\subsection*{Pearson's chi-square divergence}

We have from the previous expression
\begin{equation}\label{eq:appendKim}
  \J(\Gth) \rightarrow \sum \left(\frac{\qdot^2_i}{q_i}\right)\dd\Gth^2 
                                = \sum \frac{\dd q^2_i}{q_i}.
\end{equation}
Pearson's chi-square divergence, or chi-square test statistic, is usually described as follows.  Given an expected probability distribution, $\{q_i\}$, and an observed probability distribution, $\{q_i'\}$, the chi-square statistic is the sum of observed minus expected squared over expected. Writing the observed minus expected squared as $\GD q^2_i = (q_i'-q_i)^2$, we have
\begin{equation*}
  \chi^2(\Gth) = \sum \frac{\GD q^2_i}{q_i}.
\end{equation*}
As the changes become small, 
\begin{equation*}
  \chi^2(\Gth) \rightarrow \sum \frac{\dd q^2_i}{q_i} = \sum \left(\frac{\qdot^2_i}{q_i}\right)\dd\Gth^2,
\end{equation*}
demonstrating that the Jeffreys and chi-square divergences have the same limiting form.  The next section shows that the limiting form is related to the Fisher information metric.

When changes are large, only the Jeffreys divergence gives the correct expression for changes by selection in mean log fitness, $\GDs\mbar$.  The chi-square divergence is the change in mean fitness on a linear scale
\begin{equation*}
  \GDs\wbar = \sum \GD q_i w_i = \sum \frac{\GD q^2_i}{q_i}.
\end{equation*}

As I discussed in the text, the correct scale for analyzing changes in fitness is logarithmic, because fitness is a relative measure, and logarithmic scaling is the correct scale for relative measures \autocite{wagner10the-measurement}.  In addition, the relations between selection and information are only clear on the logarithmic scale, because it is only on that scale that one can see the connections to the classic theories of entropy and information.  In the limit of small changes, the logarithmic scale becomes linear, and thus $\GDs\mbar \rightarrow \GDs\wbar$.

\subsection*{Alternative expressions for Fisher information}

One can think of Fisher information as the change in a probability distribution with respect to a change in a parameter that specifies the distribution.  The more rapidly a distribution changes with respect to a parameter, the more information each observation provides about the value of the parameter.  For example, if the distribution changes very slowly, then small differences in the distribution of observed values may translate into big differences in parameter values.  Thus, approximately similar distributions of observations map to widely different parameter values, so each observation provides relatively little information about the parameter.  If, by contrast, the distribution changes rapidly with respect to a parameter, then the distribution of observations is very different for small changes in the parameter, and each observation provides a lot of information about the likely value of the parameter.

Mathematically, Fisher information is the negative value of the expected curvature of the log-likelihood function
\begin{equation*}
  F(\Gth) = -\sum q_i \left(\frac{\dd^2\log(q_i|\Gth)}{\dd\Gth^2}\right).
\end{equation*}
Doing the differentiation, and noting \autocite{amari00methods} that 
\begin{equation*}
  \sum\frac{\dd^2 q_i|\Gth}{\dd\Gth^2} = \frac{\dd}{\dd\Gth}\sum  \frac{\dd q_i|\Gth}{\dd\Gth}=0,
\end{equation*}
because the sum of changes in frequencies must be zero over a distribution, we obtain
\begin{equation*}
  F(\Gth) =  \sum \frac{\qdot^2_i}{q_i}.
\end{equation*}  
A large number of different divergence measures converge to Fisher information in the limit.  Thus, knowing only that the limiting form of a divergence is Fisher information only weakly constrains the associated form of divergence.  For example, from the expression above for the chi-square divergence
\begin{equation*}
  \chi^2(\Gth) \rightarrow \sum \frac{\dd q^2_i}{q_i} = \sum \left(\frac{\qdot^2_i}{q_i}\right)\dd\Gth^2,
\end{equation*}
it might be tempting, in a particular application in which Fisher information arises, to think of the chi-square divergence as somehow the natural measure of change, because the chi-square form for large changes most closely resembles the limiting Fisher information form for small changes.  In the case of selection, that conclusion would not be correct.  The Jeffreys divergence is in fact the natural measure of change, because the logarithmic scale is the natural scale for changes in fitness and for changes in information.

\section*{Appendix B: Historical aspects}

\textcite{kimura58on-the-change} noted that the change in fitness in certain models of selection is
\begin{equation}\label{eq:appendKS}
  \GDs\mbar = \sum \frac{\qdot^2_i}{q_i}.
\end{equation}
Kimura used the standard notion of change with respect to time in his study of continuous dynamics with respect to small changes.  Thus, the parameter is $\Gth\equiv t$ for time, and $\qdot = \dd q/\dd t$.  

\textcite{ewens92an-optimizing} and \textcite{edwards00fisher} provide comprehensive syntheses of the literature on the various uses of Kimura's expression, $\sum \qdot_i^2/q_i$.  The main use concerned information geometry expressions of selection dynamics on a Riemannian manifold.  Neither Ewens nor Edwards found that discussion of information geometry particularly useful.  Edwards did note that the Kimura's expression is in fact just an expression for Fisher information.  But Edwards did not think that association was useful.  

I agree with the criticisms by Ewens and Edwards within the context of how the literature had been framed.  From \textcite{kimura58on-the-change} through the various developments in the literature, the emphasis had always been on dynamics with respect to time.  I agree with Edwards that one cannot say anything very interesting about the temporal dynamics of evolutionary change from the simple expression in \Eq{appendKS} for selection.  That expression is the partial change caused by selection \autocite{price72fishers,ewens89an-interpretation,frank92fishers}, not the total evolutionary change.  The partial change gives a clear sense of what selection is doing at any moment, but provides no insight by itself about evolutionary dynamics.  

My presentation in this article is also based on Fisher information and, more generally, on the Jeffreys divergence.  Two aspects of my presentation go beyond the past work and, in my view, provide a compelling case for framing our understanding of selection in these terms.   

First, I connected selection to information theory through the general result $\GDs\mbar=\J$, the Jeffreys divergence.  This result does not depend on the limit of small changes, but instead is a general description of the nature of selection.  This result establishes the proper measure for the amount of information accumulated by selection.

Second, I related the change in information to various underlying parametric and nonparametric scales.  Those scales provide the meaning with respect to the abstract scale for encoded information that forms the basis for classical information theory.  As \textcite{edwards00fisher} emphasized, Fisher information is information about meaning with respect to underlying parameters \autocite{frank09natural}.  Earlier work implicitly used time as the parameter, which is not a meaningful way of expressing the accumulation of information.  One does not think of selection as providing information about time.  In addition to making the parametric basis for selection and information explicit, my use of the Jeffreys divergence clarified the relation of selection to classical information theory.  

Finally, I achieved greater generality than past work by respecting the fundamental distinction between selection and evolution.  Past work often tried to make general statements about evolutionary dynamics, which is not possible.  It is possible to make strong and completely general statements about the partial change caused by selection.  Such statements clarify the relations between selection and information.  One can achieve that depth and generality only by working within the fundamental limitations imposed by the distinction between selection and total evolutionary change.

I mentioned that \textcite{ewens92an-optimizing} and \textcite{edwards00fisher} concluded that past work based on the Kimura's result did not contribute significantly to understanding selection.  \textcite{ewens92an-optimizing} did develop his own extension to that theory, in which he showed an optimization principle in relation to Fisher's fundamental theorem.  \textcite{frank09natural} developed a similar idea but with a different approach that emphasized information and the Fisher information metric. Those studies derive from a partitioning of the causes of fitness, which is the topic of a future article in this series.

\end{document}